\documentclass[12pt]{article}
\usepackage{epsfig}

\date{}
\begin{document}

\newcommand{\beq}{\begin{equation}}
\newcommand{\eeq}{\end{equation}}
\newcommand{\nn}{\nonumber}
\newcommand{\bea}{\begin{eqnarray}}
\newcommand{\eea}{\end{eqnarray}}

\title{Dark Radiation and Localization of Gravity on the
  Brane}

\author{Rui Neves\footnote{E-mail: \tt rneves@ualg.pt}\hspace{0.2cm} and Cenalo Vaz\footnote{E-mail: \tt cvaz@ualg.pt}\\
{\small \em \'Area Departamental de F\'{\i}sica/CENTRA, FCT,
Universidade do Algarve}\\
{\small \em Campus de Gambelas, 8000-117 Faro, Portugal}
}

\maketitle

\begin{abstract}
We discuss the dynamics of a spherically symmetric dark
radiation vaccum in the Randall-Sundrum brane world scenario. 
Under certain natural assumptions we show that the Einstein equations 
on the brane form a closed system. For a de Sitter
brane we determine exact dynamical and inhomogeneous solutions which
depend on the brane cosmological constant, on the dark radiation tidal 
charge and on its initial configuration. We define the conditions
leading to singular or globally regular solutions.
We also analyse the localization of gravity near the
brane and show that a phase transition to a regime where gravity
propagates away from the brane may occur at short distances during the
collapse of positive dark energy density.
\end{abstract}

\section{Introduction}

In the search for extra spatial dimensions the
Randall and Sundrum (RS) brane world scenario is particularly
interesting for its simplicity and depth \cite{RS}. In this model the 
Universe is a 3-brane boundary of a noncompact $Z_2$ symmetric  
5-dimensional anti-de Sitter space. The matter fields live only on 
the brane but gravity inhabits the whole bulk and is localized near the
brane by the warp of the infinite fifth dimension.

Since its discovery many studies have been done within the RS
scenario (see Ref.~\cite{RM1} for a recent review and notation). 
For a brane bound observer \cite{SMS,SSM,RM2} the interaction between 
the brane and the bulk introduces correction terms to the 
4-dimensional Einstein equations,
namely, a local high energy embedding term generated by the matter
energy-momentum tensor and a non-local term induced
by the bulk Weyl tensor. Such equations have an intrincate non-linear 
dynamics. For example, the exterior vaccum of collapsing matter on the
brane is now filled with gravitational modes
originated by the bulk Weyl curvature and can no longer be regarded as
a static space \cite{BGM,GD}.

Previous research on the RS scenario has been focused on static 
or homogeneous dynamical solutions. In this proceedings we report 
some new results on the dynamics of a spherically symmetric RS brane 
world vaccum. For a de Sitter brane we present exact
dynamical and inhomogeneous solutions, define the conditions to
characterize them as singular or globally regular and 
discuss the localization of gravity to the vicinity of the 
brane (see Ref.~\cite{RC} for more details).    

\section{Brane Vaccum Field Equations}

In the Gauss-Codazzi formulation of the RS model \cite{SMS,SSM,RM2},
the Einstein vaccum field equations on the brane are given by

\begin{equation}
{G_{\mu\nu}}=-\Lambda{g_{\mu\nu}}-{{\mathcal{E}}_{\mu\nu}},\label{4defe}
\end{equation}                 
where $\Lambda$ is the brane cosmological constant and 
the tensor ${{\mathcal{E}}_{\mu\nu}}$ is the limit on the
brane of the projected 5-dimensional Weyl tensor. It is a symmetric and
traceless tensor constrained by the following conservation equations

\begin{equation}
{\nabla_\mu}{{\mathcal{E}}^\mu_\nu}=0\label{ce}.
\end{equation}

The projected Weyl tensor ${{\mathcal{E}}_{\mu\nu}}$ 
can be written in the following general form \cite{RM2}

\begin{equation}
{{\mathcal{E}}_{\mu\nu}}=-{{\left({{\tilde{\kappa}}\over{\kappa}}\right)}^4}
\left[{\mathcal{U}}
\left({u_\mu}{u_\nu}+{1\over{3}}{h_{\mu\nu}}\right)
+{{\mathcal{P}}_{\mu\nu}}+{{\mathcal{Q}}_\mu}{u_\nu}+
{{\mathcal{Q}}_\nu}{u_\mu}\right],
\end{equation}
where $u_\mu$ such that ${u^\mu}{u_\mu}=-1$ is the 4-velocity field
and ${h_{\mu\nu}}={g_{\mu\nu}}+{u_\mu}{u_\nu}$ is the tensor
projecting orthogonaly to
$u_\mu$. The forms $\mathcal{U}$, ${\mathcal{P}}_{\mu\nu}$ and 
${\mathcal{Q}}_\mu$ represent different aspects of the effects induced
on the brane by the 5-dimensional gravitational field. Thus,
$\mathcal{U}$ is an energy density,
${\mathcal{P}}_{\mu\nu}$ a stress tensor and ${\mathcal{Q}}_\mu$ an
energy flux. 

Since the 5-dimensional metric is not known, in general
${\mathcal{E}}_{\mu\nu}$ is not completely determined on the brane
\cite{SMS,SSM} and so the effective 4-dimensional theory is not
closed. To close it we need simplifying assumptions about the effects
of the gravitational field on the brane. For instance we may consider 
a static and spherically symmetric brane vaccum with ${{\mathcal{Q}}_\mu}=0$, 
${{\mathcal{P}}_{\mu\nu}}\not=0$ and $\mathcal{U}\not=0$. This leads
to the Reissner-Nordstr\"om black hole solution on the brane \cite{DMPR}.    

It is also possible to close the system of Einstein equations when 
considering a dynamical and spherically symmetric
brane vaccum with ${{\mathcal{Q}}_\mu}=0$, $\mathcal{U}\not=0$,
and ${{\mathcal{P}}_{\mu\nu}}\not=0$. The general,
spherically symmetric metric in comoving coordinates
$(t,r,\theta,\phi)$ is given by

\begin{equation}
d{s^2}={g_{\mu\nu}}
d{x^\mu}d{x^\nu}=-{e^\sigma}d{t^2}+A^2d{r^2}+{R^2}d{\Omega^2},
\label{met}
\end{equation}
where $d{\Omega^2}=d{\theta^2}+{\sin^2}
\theta d{\phi^2}$, $\sigma=\sigma(t,r)$, $A=A(t,r)$, $R=R(t,r)$ and
$R$ is the physical spacetime radius. If the traceless stress tensor 
${\mathcal P}_{\mu\nu}$ is isotropic then it will have the general form 

\begin{equation} 
{{\mathcal P}_{\mu\nu}}={\mathcal{P}}\left({r_\mu}{r_\nu}-{1\over{3}}{h_{\mu\nu}}\right),
\end{equation}
where ${\mathcal{P}}={\mathcal{P}}(t,r)$ and $r_\mu$ is the unit radial vector, given in the
above metric by ${r_\mu}=(0,A,0,0)$. Then

\begin{equation}
{{\mathcal{E}}_\mu^\nu}={{\left({{\tilde{\kappa}}\over{\kappa}}\right)}^4}\mbox{diag}\left(\rho,-
{p_r},-{p_T},-{p_T}\right),
\end{equation}
where the energy density and pressures are, respectively, 
$\rho={\mathcal{U}}$,
${p_r}=(1/3)\left({\mathcal{U}}+2{\mathcal{P}}\right)$ and ${p_T}=(1/3)\left({\mathcal{U}}-{\mathcal{P}}
\right)$. Consequently, the conservation Eq.~(\ref{ce}) read \cite{TPS}

\[
2\frac{\dot A}{A}\left(\rho+{p_r}\right)=-2\dot{\rho}-
4{{\dot{R}}\over{R}}\left(\rho+{p_T}\right),
\]
\begin{equation}
\sigma'\left(\rho+{p_r}\right)=-2{p_r}'+4{{R'}\over{R}}({p_T}-{p_r}),\label{cee}
\end{equation}
where the dot and the prime denote, respectively, derivatives
with respect to $t$ and $r$. A synchronous solution is permitted with
the equation of state $\rho+{p_r}=0$, equivalent to 
${\mathcal{P}}=-2{\mathcal{U}}$ where $\mathcal{U}$ has 
the dark radiation form 

\begin{equation}
{\mathcal{U}}={{\left({{\kappa}\over{\tilde{\kappa}}}\right)}^4}
{Q\over{R^4}}.\label{KfB}
\end{equation}
The constant $Q$ is the dark radiation tidal charge. Hence, we get

\begin{equation}
{G_{\mu\nu}}=-\Lambda{g_{\mu\nu}}+{Q\over{R^4}}\left({u_\mu}{u_\nu}-
2{r_\mu}{r_\nu}+{h_{\mu\nu}}\right),\label{dem}
\end{equation}
an exactly solvable closed system for the unknown functions $A(t,r)$ and $R(t,r)$ which depends on the free parameters $\Lambda$ and $Q$. Indeed, its
solutions are of the LeMa\^{\i}tre-Tolman-Bondi type  

\begin{equation}
d{s^2}=-d{t^2}+{{{R'}^2}\over{1+f}}d{r^2}+{R^2}d{\Omega^2},
\end{equation}
where $R$ satisfies

\begin{equation}
\dot{R}^2={\Lambda\over{3}}{R^2}-{Q\over{R^2}}+f.
\label{deq}
\end{equation}
The function $f=f(r)>-1$ is interpreted as the energy inside 
a shell labelled by $r$ in the dark radiation vaccum and is fixed
by its initial configuration.

\section{Localization of Gravity near the Brane}

As is clear in Eq.~(\ref{dem}) the dark radiation dynamics depends on 
$\Lambda$ and $Q$. It is important to point out that these parameters
have a direct effect on the localization of gravity in the vicinity 
of the brane. Indeed, the tidal acceleration away from the brane
\cite{RM2} is given by \cite{DMPR}

\begin{equation}
-{\lim_{y\to 0\pm}}{\tilde{R}_{ABCD}}{n^A}{\tilde{u}^B}{n^C}{\tilde{u}^D}=
{{\tilde{\kappa}^2}\over{6}}\tilde{\Lambda}+{Q\over{R^4}},
\end{equation}
where $\tilde{u}_A$ is the
extension off the brane of the 4-velocity field satisfying
${\tilde{u}^A}{n_A}=0$ and ${\tilde{u}^A}{\tilde{u}_A}=-1$. The
gravitational field is only bound to the brane if the tidal
acceleration points towards the brane. It must then be negative
implying that

\begin{equation}
{\tilde{\Lambda}}{R^4}<-{{6Q}\over{\tilde{\kappa}^2}}.
\end{equation}     
As a consequence, gravity is only localized for all $R$ if 
$\Lambda<{\Lambda_c}$ with ${\Lambda_c}=
{\tilde{\kappa}^4}{\lambda^2}/12$ and
$Q\leq 0$ or $\Lambda={\Lambda_c}$ and $Q<0$. For
$\Lambda<{\Lambda_c}$ and $Q>0$ the
gravitational field will just remain localized if $R>{R_c}$ where ${R_c^4}=3Q/({\Lambda_c}-\Lambda)$. On the
other hand for $\Lambda>{\Lambda_c}$ and
$Q<0$ localization is limitted to the epochs $R<{R_c}$. If
$\Lambda\geq{\Lambda_c}$ and $Q\geq 0$ then
gravity is always free to propagate far away into the bulk.

According to recent supernovae measurements (see {\it{e.g.}}
Ref.~\cite{SND}) $\Lambda\sim{10^{-84}}{\mbox{GeV}^2}$. On the other hand
${\tilde{M}_{\mbox{\footnotesize p}}}>{10^8}\mbox{GeV}$ and
${M_{\mbox{\footnotesize p}}}\sim{10^{19}}\mbox{GeV}$ imply 
$\lambda>{10^8}{\mbox{GeV}^4}$ \cite{LMSW} because 
$6{\kappa^2}=\lambda{\tilde{\kappa}^4}$. Since 
${\Lambda_c}={\kappa^2}\lambda/2$ then ${\Lambda_c}$ is bound from
below, ${\Lambda_c}>{10^{-29}}{\mbox{GeV}^2}$. Hence, observations
demand $\Lambda$ to be positive and smaller than the critical
value $\Lambda_c$, $0<\Lambda<{\Lambda_c}$. Note that is means an
anti-de Sitter bulk, $\tilde{\Lambda}<0$. The same conclusion is true
if $M_{\mbox{\footnotesize
      p}}$ is in the TeV range because $\Lambda_c$ increases when
$M_{\mbox{\footnotesize p}}$ decreases.  

Since current observations do not yet constrain the sign of $Q$ \cite{qexp} we
conclude that for $0<\Lambda<{\Lambda_c}$ only for $Q<0$ gravity is
bound to the brane for all $R$. If $Q>0$ then for $R<{R_c}$ the tidal
acceleration is positive and gravity is no longer localized near the brane.

\section{Inhomogeneous Dynamics for a de Sitter Brane}

Assume from now on that $0<\Lambda<{\Lambda_c}$. Non-static solutions
correspond to $f\not=0$. An example is  

\begin{equation}
\left|{R^2}+{{3f}\over{2\Lambda}}\right|=\sqrt{\beta}\cosh\left[\pm 2
\sqrt{{\Lambda\over{3}}}t+{\cosh^{-1}}\left(
{{\left|{r^2}+{{3f}\over{2\Lambda}}\right|}
\over{\sqrt{\beta}}}\right)\right],\label{sol1}
\end{equation}
where $\beta=(3/\Lambda)[3{f^2}/(4\Lambda)+Q]$ and $+$ or $-$
correspond respectively to expansion or collapse. If $Q>0$
then $f>-1$ but for $Q<0$
the energy function $f$ must satisfy in addition
$|f|>2\sqrt{-Q\Lambda/3}$. 
Since $R$ is a non-factorizable function of $t$ and $r$ these
solutions define new exact and inhomogeneous cosmologies for the
spherically symmetric dark radiation de Sitter brane. 

\section{Singularities and Regular Bounces}

The dark radiation dynamics defined by Eq.~(\ref{deq})
may produce shell focusing singularities at
$R=0$ or regular bouncing points at some $R\not=0$. To see this
consider

\begin{equation}
{R^2}{\dot{R}^2}=V(R,r)={{\Lambda}\over{3}}{R^4}+f{R^2}-Q.
\end{equation}    
If for all $R\geq 0$ the potential $V$ is positive then a shell
focusing singularity forms at $R={R_s}=0$. Alternatively, if there is
an epoch $R={R_*}\not=0$ for which $V=0$ then a regular rebounce point
appears at $R={R_*}$. For the dark radiation vaccum at most two
regular rebounce epochs can be found. Since $\Lambda>0$ there is always a phase of continuous
expansion to infinity with ever increasing rate. Depending on $f(r)$ 
other phases may exist. To ilustrate
take $\beta>0$ and compare the settings
$Q<0,f>-1,|f|>2\sqrt{-Q\Lambda/3}$ and $Q>0,f>-1$. 

\begin{figure}[ht]
\setlength{\unitlength}{1cm}
\centerline{\epsfxsize=7cm\epsfysize=4cm\epsfbox{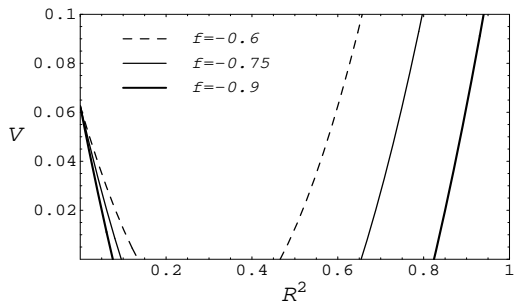}}
\caption{\small Plot of $V$ for $\beta>0$ and $Q<0$. Non-zero values of
  $f$ belong to the
interval $-1<f<-2\sqrt{-Q\Lambda/3}$ and correspond to
shells of constant $r$.\label{fig:Vppn}}
\end{figure}
If for $Q<0$ we have 
$f>2\sqrt{-Q\Lambda/3}$ then $V>0$ for all $R\geq 0$. 
There are no rebounce points 
and the dark radiation shells may either expand
continuously or collapse to a singularity at ${R_s}=0$.
However for $-1<f<-2\sqrt{-Q\Lambda/3}$ (see Fig.~\ref{fig:Vppn}) 
we find two rebounce
epochs at $R=R_{*\pm}$ with ${R_{*\pm}^2}=-3f/(2\Lambda)\pm\sqrt{\beta}$.
Since $V(0,r)=-Q>0$ a singularity also forms at ${R_s}=0$. Between the 
two rebounce points there is a forbidden zone where $V$ 
is negative. The phase space of allowed dynamics is thus divided in 
two disconnected regions separated by the forbidden interval 
${R_{*-}}<R<{R_{*+}}$. For $0\leq R\leq{R_{*-}}$ the
dark radiation shells may expand to a maximum radius $R={R_{*-}}$, 
rebounce and then fall to the singularity.
If $R\geq{R_{*+}}$ then there
is a collapsing phase to the minimum radius $R={R_{*+}}$ 
followed by reversal and subsequent accelerated continuous expansion. 
The singularity at ${R_s}=0$ does not form and so the solutions are globally
regular. Since $Q<0$ gravity is bound to the brane for all the values
of $R$.

\begin{figure}[ht]
\setlength{\unitlength}{1cm}
\centerline{\epsfxsize=7cm\epsfysize=4cm\epsfbox{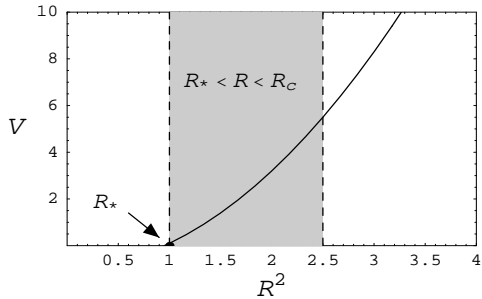}}
\caption{\small Plot of $V$ for $\beta>0$ and $Q>0$. Non-zero values of
  $f$ belong to the
interval $f>-1$ and correspond to
shells of constant $r$. The shaded region indicates where gravity is
not localized near the brane.\label{fig:Vppp}}
\end{figure}
\vspace{-0.5cm}
If $Q>0$ (see Fig.~\ref{fig:Vppp}) then we find globally
regular solutions with a single rebounce epoch at $R={R_*}$ where
${R_*^2}=-3f/(2\Lambda)+\sqrt{\beta}$. This is the minimum possible 
radius for a collapsing dark radiation shell. It then reverses its
motion and expands forever. The phase
space of allowed dynamics defined by $V$ and $R$ is limitted to the
region $R\geq{R_*}$. Below
$R_*$ we find a forbidden region where $V$ is negative.
In particular, $V(0,r)=-Q<0$ implying that the
singularity at ${R_s}=0$ does not form and so the solutions are globally
regular. Note that if gravity is to
be bound to the the brane for $R>{R_*}$ then
${R_*}>{R_c}$. If not then
we find a phase transition epoch $R={R_c}$ such that for $R\leq{R_c}$
the gravitational field is no longer localized near the brane.

\section{Conclusions}

In this work we have reported some new results on the dynamics 
of a RS brane world dark radiation vaccum. Using an 
effective 4-dimensional approach we have shown that some simplifying
but natural assumptions lead to a closed and solvable system of 
Einstein field equations on the brane. We have presented a set of
exact dynamical and
inhomogeneous solutions for $\Lambda>0$ showing they
further depend on the dark radiation tidal charge $Q$ and on the energy 
function $f(r)$. We have also described the 
conditions under which a singularity or a regular rebounce point develop
inside the dark radiation vaccum and discussed the localization of
gravity near the brane. In particular, we have shown that a phase 
transition to a regime where gravity is
not bound to the brane may occur at short distances during the
collapse of positive dark energy density on a realistic de Sitter
brane. Left for future research is
for example an analysis of the dark radiation vaccum dynamics from a
5-dimensional perspective.

\section*{Acknowledgements}
We thank the Funda\c {c}\~ao para a Ci\^encia e a Tecnologia (FCT) for
financial support under the contracts POCTI/SFRH/BPD/7182/2001 and 
POCTI/32694/FIS/2000. We also would like to thank Louis Witten and
T.P. Singh for kind and helpful comments.

\end{document}